\documentclass[reprint,amsmath,amssymb,amsfonts,aps,prb,superscriptaddress]{revtex4-2}
\usepackage{graphicx}
\usepackage{bm}
\usepackage{epsfig}
\usepackage{epstopdf}
\usepackage{xcolor}
\usepackage{hyperref}
\hypersetup{colorlinks=true,allcolors=blue}
\usepackage{natbib}
\usepackage{mathtools}
\usepackage{siunitx}
\usepackage{soul,xcolor}
\usepackage{amsmath,amssymb}

\setstcolor{red}
\DeclareUnicodeCharacter{2212}{-}
\DeclarePairedDelimiterXPP\BigOSI[2]%
  {\mathcal{O}}{(}{)}{}%
  {\SI{#1}{#2}}

\begin{document}
\title{Global Quench Dynamics and the Growth of  Entanglement Entropy in Disordered Spin Chains with Tunable Range Interactions}

\author{Y. Mohdeb}
\email[]{y.mohdeb@jacobs-university.de}
\affiliation{Department of Physics and Earth Sciences, Constructor University, formerly Jacobs University Bremen, Bremen 28759, Germany}  

\author{J. Vahedi}
\email[]{javahedi@kip.uni-heidelberg.de} 
\affiliation{Department of Physics and Earth Sciences, Constructor University, formerly Jacobs University Bremen, Bremen 28759, Germany} 
\affiliation{Kirchhoff-Institut für Physik, Universität Heidelberg, Im Neuenheimer Feld 227, 69120 Heidelberg, Germany}
\affiliation{Department of Physics, Sari Branch, Islamic Azad University, Sari 48164-194, Iran}

\author{R. N. Bhatt}
\email[]{ravin@princeton.edu}
\affiliation{Department of Electrical Engineering, Princeton University, Princeton, New Jersey 08544, USA}
\author{S. Haas}
\email[]{shaas@usc.edu}
\affiliation{Department of Physics and Astronomy University of Southern California, Los Angeles, CA 90089-0484}
\affiliation{Department of Physics and Earth Sciences, Constructor University, formerly Jacobs University Bremen, Bremen 28759, Germany}
\author{S. Kettemann}
\email[]{s.kettemann@jacobs-university.de}
\affiliation{Department of Physics and Earth Sciences, Constructor University, formerly Jacobs University Bremen, Bremen 28759, Germany}
\affiliation{Division of Advanced Materials Science, Pohang University of Science and Technology (POSTECH), Pohang 790-784, South Korea}

\begin{abstract} 
   The non-equilibrium dynamics of disordered  many-body  quantum systems after a global quantum quench  unveils  important insights about the competition between interactions and disorder, yielding in particular  an insightful perspective on 
   many body localization (MBL). Still, the experimentally relevant  effect of  bond randomness in long-range interacting  spin chains on the quantum quench dynamics have so far not been investigated.
In this letter, we examine  the entanglement entropy growth after a global  quench in a  quantum spin chain with randomly placed spins and long-range tunable interactions decaying with distance with  power $\alpha$. Using a dynamical version of the strong disorder renormalization group (SDRG)  we find  for $\alpha >\alpha_c$ that  the entanglement entropy grows logarithmically with time  and becomes smaller  with larger $\alpha$ as  $S(t) = S_p  \ln(t)/(2\alpha)$. Here, $S_p= 2 \ln2 -1$.  We use  numerical exact diagonalization (ED) simulations to verify our results for system sizes up to $ N\sim 16$ spins, yielding  good agreement  for sufficiently large $\alpha > \alpha_c \approx 1.8$. For $\alpha<\alpha_c$, we find that the entanglement entropy grows as a power-law with time, $S(t)\sim t^{\gamma(\alpha)}$ with $0<\gamma(\alpha)<1$ a decaying function of the interaction exponent $\alpha$.
\end{abstract} 

\maketitle
Magnetic resonance experiments in doped semiconductors\cite{Feher1955} motivated P. W. Anderson to address the issue of electron localization in disordered systems using a model of noninteracting electrons \cite{anderson58}. Later, Fleischman and Anderson \cite{Fleishman1980} argued that short-range electron-electron interactions with localized single-particle states would preserve the basic results of
Anderson’s non-interacting model, namely the existence of a localized phase for strong disorder. The issue of localization in a many-body system was put on more rigorous footing in
references\cite{basko06,gornyi05} using a perturbative approach, where the concept of many-body localization
(MBL) was introduced. Since then, the study of MBL in disordered interacting systems, has
become a flourishing field; for recent reviews, see \cite{Abanin2019,Bhatt2021} and references therein. In the presence of  interactions with a power-law dependence on distance, it is well established that a many-body localized phase  persists in random spin chains as long as the interactions fall-off faster than a certain critical power law\cite{burin15,Burin2015,Schiffer2019,Nandkishore2017}. 
 On the other hand, a logarithmic divergence of the  entanglement entropy with subsystem size
 $n$ at finite energy density  as $S_A \sim ln(n)$ and   average correlations decaying with a power law was found to occur in models with bond-disorder and particle-hole symmetry\cite{Vasseur2015,Vasseur2016}. This phase was  dubbed \begin{it}quantum critical glass\end{it} (QCG) and found to survive the introduction of long-range interactions provided that its power-law decay has an exponent which exceeds a critical value $\alpha_c$\cite{Mohdeb2022}. The full characterization of (marginally) localized phases in long-range interacting random systems remains  greatly unexplored  and several questions have yet to be answered. While most  studies   focused  so far on disorder in the form of random potential with the aim of characterizing MBL, disorder in the  long-range  interactions is ubiquitous in real quantum systems\cite{anderson58,mott76,Signoles2021}.

An insightful perspective of the  delocalization-localization transitions in random lattice spin models is provided by the entanglement entropy  (EE) dynamics after a quantum quench. For global quenches, the system is usually prepared in a nonentangled initial state corresponding to the ground state of a known Hamiltonian, and then let evolve in time with the initial Hamiltonian.
This approach has been widely used as a probe of many-body localization for both short-range interacting archetypal random models\cite{nidari2008,bardarson12,Serbyn2013}, and long-range interacting spin chains with random local magnetic fields\cite{pino14,SafaviNaini2019,DeTomasi2019}. For MBL systems with nearest neighbor interactions, it has been  shown\cite{nidari2008,bardarson12,Serbyn2013}  that the EE grows logarithmically with time after a quantum quench from an unentangled high-energy state $S(t)\sim \ln(t)$  until reaching a saturation value which is determined by the participation ratios of the initial state over the eigenstates of the subsystem\cite{Serbyn2013}. In the presence of long-range interactions some studies found that the EE grows as  a power-law with time $S(t)\sim t^{1/\alpha}$\cite{SafaviNaini2019,pino14}. Recently, it has been suggested that  at  the MBL transition  in long-range interacting spin models  subject to random magnetic fields the EE grows also with a power law in time, albeit with a universal exponent  $\delta\approx 0.33$\cite{Deng2020}.

It is therefore of great interest to study the entanglement dynamics in spin chains with long range interactions in which randomness is present in the interactions themselves. The random bond XX-spin chain with nearest neighbor interactions is known to be a quantum critical glass (QCG)\cite{Vasseur2016}.
 There, a strong disorder renormalization group procedure\cite{dasguptama,bhatt82, Bhatt2021,monthus,vosk13,fisher94} has been applied to study the dynamics of the EE\cite{vosk13}, and more recently of  Rényi entropies\cite{Ruggiero2022}. An  ultra-slow dynamics was found where the EE scales as $S(t)\sim \ln( \ln (t))$.  
 How random bond 
  long-range interactions affect the entanglement  dynamics is the topic of this letter.

\begin{it}
Model.\end{it}— In this letter, we  employ  a combination of  dynamical strong disorder renormalization group (SDRG) approach, also known as RSRG-t\cite{vosk13,Vosk2014,Monthusflo}, and numerical exact diagonalization (ED) to investigate the dynamics of the entanglement entropy  after a global quantum quench in a long-range tunable interacting XX-spin chain with positional disorder. 
 We consider the Hamiltonian:
\begin{equation}\label{H}
H=\sum_{i<j}J_{ij}\left(S_{i}^{x}\,S_{j}^{x}+S_{i}^{y}\,S_{j}^{y}\right) 
\end{equation}
describing $N$ interacting $S=1/2$ spins that are randomly placed at positions ${\bf r}_i$ on a  lattice of length $L$ and lattice spacing  $a$, with density $n = N/L$.  The couplings between all pairs of sites $i,j,$ are taken to be antiferromagnetic and long-ranged, decaying with a power law, 
\begin{equation} \label{jcutoff}
J_{ij} = J_0\left|({\bf r}_i-{\bf r}_j)/a\right |^{-\alpha}.
\end{equation}
We consider open boundary conditions. 
The entanglement  properties of this model were previously investigated for both  the ground state and  generic excited eigenstates by means of SDRG and ED in Refs. \cite{Mohdeb2020,Mohdeb2022}.
It was found that  the ground state of the model is correctly captured by a random singlet phase, with a 
 distribution of 
couplings which   flows to a strong disorder fixed point (SDFP), as characterized by a finite dynamical exponent $z=2\alpha$. More recently,  the  eigenstates in the middle of the many-body spectrum of this model were studied\cite{Mohdeb2022}.
A  delocalized regime was found  for $\alpha \leq \alpha_c$, characterized by an algebraic sub-volume enhancement of the entanglement entropy with subsystem size.  For $\alpha \geq \alpha_c\approx 1$ the infinite temperature eigenstates were found to be marginally localized; a logarithmic scaling of the entanglement entropy with subsystem size $S_n\sim \ln(n)$ was found, indicating that the system is in a QCG phase for sufficiently large $\alpha$.
  
\begin{it}
SDRG.—
\end{it}
Let us recall how to  apply the SDRG to the model  Eq. (\ref{H}).
Choosing the pair $(i,j)$ with  the largest coupling $J_{i,j}$, which 
forms a singlet, we     take the expectation value of the Hamiltonian 
        in that particular singlet state within second-order perturbation  theory
             in  the couplings with
              all other spins.
              For long range interactions this yields the  
           renormalization rule
             for the  couplings between  spins $(l,m)$ in the XX model as given by \cite{Mohdeb2020,ourPRB}
              \begin{equation} \label{jeff}
               (J^{x}_{lm})' =   J_{lm}^x - \frac{(J^x_{li}-J^x_{lj})(J^x_{im}-J^x_{jm})}{J^x_{ij}}.
              \end{equation}
 In the short-range case,  these
 RG equations lead to an infinite randomness fixed point (IRFP), where the distribution of renormalized  couplings  gets wider at every RG step,
  having a 
  width $W=(\langle \ln (J/\Omega_0)^2 \rangle -  \langle \ln (J/\Omega_0) \rangle^2)^{1/2} =\ln(\Omega_0/\Omega) = \Gamma_{\Omega},$ 
  which increases monotonically as  the RG scale $\Omega$ is lowered. In contrast, for long-range couplings with 
 finite $\alpha$ the width $W$  saturates and   converges to  $W = \Gamma$, with $\Gamma = 2 \alpha$ for the XX model, characterizing the strong disorder fixed point (SDFP)\cite{ourPRB}.
For large number of spins $N \gg 1,$  and in the limit of  small RG scale $\Omega$,     the resulting 
     distribution function of renormalized couplings $J$ at RG scale $\Omega$
     was at the SDFP found to converge 
      to\cite{monthus},
      \begin{equation} \label{pjsd}
      P(J,\Omega) =\frac{1}{\Omega \Gamma_{\Omega}} \left( \frac{\Omega}{J}\right)^{1-1/\Gamma_{\Omega}}.
      \end{equation}
  At the IRFP, $\Gamma_{\Omega}$ increases monotonically as 
  $\Gamma_{\Omega}= \ln \Omega_0/\Omega$,  
  when $\Omega$, the largest energy at this renormalization step, is lowered. Here 
   $\Omega_0$ is the initially largest energy in the spin chain.
  At the  SDFP, however,   $\Gamma_{\Omega}$ is found to  converge to a finite value $\Gamma_{\Omega} \rightarrow  2 \alpha$\cite{Mohdeb2020,ourPRB}, yielding
   a  distribution with finite width $W=\Gamma$.

\begin{it}
RSRG-t.—
\end{it}
The time-dependent real-space renormalization group (RSRG-t)
is an extension of the SDRG to nonequilibrium setups.
RSRG-t is designed to construct the
effective dynamics via the iterative elimination of
 degrees of freedom which oscillate with the highest frequency $\Omega$. Thereby, the RG decimation does not project the spin pairs into singlet states, as in the SDRG case,  but rather generates effective degrees of freedom which define the late-time dynamics of the system\cite{Vosk2014,vosk13}.
RSRG-t  thereby yields an effective time-independent Hamiltonian $H_{eff}$, via successive elimination of the fastest oscillating pair of spins on sites
$i$ and $j$, coupled by $J_{ij}=\Omega$, with $\Omega=\max\{ J_{ij}\}$ which dominate the  short-time dynamics. 
In presence of strong disorder,  the frequency of the eigenmodes of the largest term in the Hamiltonian, is much larger than those of the undecimated spins. Hence, the sites $i$ and $j$ are  seen by the remaining degrees of freedom as in a time-averaged state. The remaining degrees of freedom  can then be treated perturbatively.

In Ref.\cite{Monthusflo} the equivalence between this approach and the RSRG-X, which is an extension of the SDRG to excited eigenstates\cite{Pekker2014}, was outlined and derived in the framework of Floquet high frequency expansion. 
We intend to apply this procedure to the Hamiltonian Eq. (\ref{H}). 
Following\cite{Monthusflo}, we define the projectors associated to the spins $(i,j)$ as  $P_{\mu}=|\mu\rangle\langle\mu|$, with $\mu=1, 2, 3, 4$ and 
$|1\rangle=|\hspace{-.3em}\uparrow \uparrow\rangle$, 
$|2\rangle=|\hspace{-.3em}\downarrow \downarrow\rangle$, $|3\rangle=2^{-1/2} \left(|\uparrow\,\downarrow\rangle+|\downarrow\,\uparrow\rangle\right)$,
$|4\rangle=2^{-1/2} \left(|\uparrow\,\downarrow\rangle-|\downarrow\,\uparrow\rangle\right)$.

In second order perturbation theory we find  that the couplings are renormalized as
\begin{equation}
\begin{split}
&(J_{lm})_r= J_{lm}-  \frac{J_{i l} J_{jm} +J_{im} J_{jl} }{J_{ij}} (P_1+ P_2) +\\
 &\frac{(J_{li}+J_{lj})(J_{im}+J_{jm})}{J_{ij}}  P_3 -  \frac{(J_{li}-J_{lj})(J_{im}-J_{jm})}{J_{ij}}P_4.
\end{split}
\end{equation}
These RG-rules correspond to the result found using RSRG-X on the same model\cite{Mohdeb2022}. This is not surprising, since  the derivation of the effective Hamiltonian is equivalent\cite{Monthusflo}. However, when describing the dynamics, the interpretation is different   as outlined above.
\begin{figure}
    \includegraphics[scale=0.5]{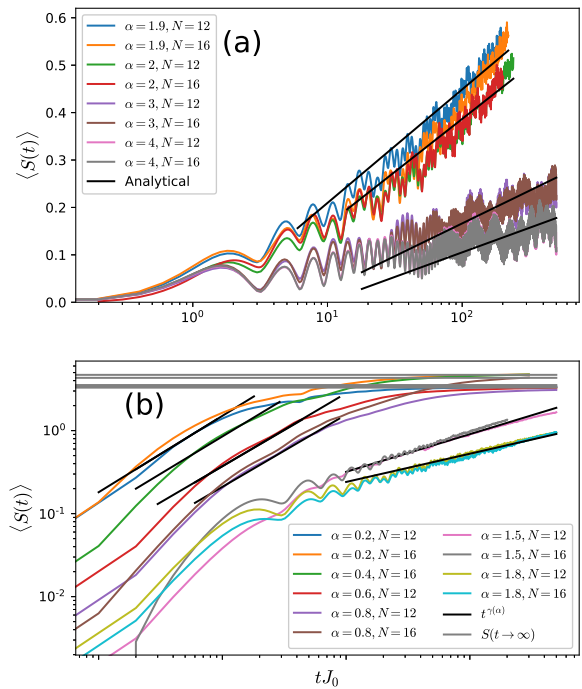}
\vspace*{-1cm}\caption{(a) Half-chain EE as function of time for different values of $\alpha> \alpha_c\approx 1.8$  obtained via ED and compared with
the analytical 
RSRG-t result Eq.(\ref{Sdyn}) for $N=12$, $N=16$ spins, $n=N/L=0.1$ and up to $1000$ disorder realizations. (b) Half chain EE as a function of time for $\alpha\leq\alpha_c$. After an initial transient time, the EE is found to grow as $S(t) \sim t^{\gamma(\alpha)}$ indicated by the solid black lines, reaching a limiting value  $S(t\rightarrow\infty)$, Eq. (\ref{slimit}).  For $\alpha=0.2,0.4,0.6, 0.8, 1.5, 1.8$ (grey lines), we find   $\gamma=0.92,0.9, 0.88, 0.86, 0.45 \hspace*{0.1cm} \text{and}\hspace*{0.1cm} 0.34$,  respectively. Here $J_0=1$ was considered.} 
    \label{DynEE}
\end{figure}

\begin{it}
Entanglement entropy.—
\end{it}
The entanglement dynamics is monitored by means of the EE $S(t)= -\text{Tr}(\rho_A \ln(\rho_A)$, where $\rho_A =\text{Tr}_B(\rho)$ is  the reduced density matrix of the  the subsystem (A) when tracing over its complement (B).
The knowledge of the fixed-point distribution of the couplings within the RSRG-t flow gives direct access to the entanglement entropy growth with time.
Given an arbitrary bipartition of an infinite spin chain, the entanglement between the 2 parts of the bipartition is then due to the  oscillating
pairs connecting the two parts of the system, forming up to an RG scale $\Omega \sim \frac{1}{t}$. In Ref. \cite{Mohdeb2022} we  have shown that for $\alpha \geq \alpha_c$ the same SDFP distribution as in the ground state, Eq. (\ref{pjsd}) is obtained within the RSRG-X flow for excited eigenstates. To find $S(t)$, it is thus sufficient to derive  the number of singlets and entangled  triplets, forming over a  bond at RG-time $\Omega \sim 1/t$, $n_{\Omega}$, as given by \cite{refael04},
 \begin{equation} 
\label{Numb}
    dn_{\Omega}= P(J=\Omega,\Omega) d\Omega.
\end{equation}
For $\Omega\rightarrow 0$ the SDFP distribution is given by Eq. (\ref{pjsd}), yielding 
$n_{\Omega} =  1/(2\alpha) \ln(\Omega)$.
Since the only  entanglement generating mechanism is  the decimation of pairs,
whose spins reside on opposite sides of the interface, one finally obtains with $\Omega\sim 1/t$ that  $S(t) = S_p \frac{1}{2\alpha} \ln(t) $, where $S_p$ is the time-averaged entanglement contribution of a decimated pair of spins, which is found to be $ S_p= 2\ln2- 1 $\cite{vosk13}.
Thus, we find
\begin{equation}
\label{Sdyn}
    S(t) = (2\ln2-1) \frac{1}{2\alpha} \ln(t).
\end{equation}
Note that the obtained prefactor is specific to $U(1)$ symmetric initial states such as the Néel state\cite{vosk13}, where only the singlet and the entangled triplet states are populated within the RSRG-t flow, which   contribute to the entanglement entropy equally. For other initial state, the logarithmic growth of the EE for $\alpha>\alpha_c$ is expected to still hold, albeit with a different prefactor.

The obtained logarithmic growth of 
 entanglement entropy  Eq. (\ref{Sdyn})  is faster than  the one obtained for the nearest neighbour XX spin chain with random bonds after a global quench, $S(t) \sim\ln(\ln(t))$\cite{vosk13}. A  logarithmic increase with time is  known rather to occur in  conventional MBL  with short-range interacting systems and random potential\cite{bardarson12,nidari2008}.
In Ref. \cite{Moessner2017}
a logarithmic growth of EE has been obtained for a model of Fermions with long range hoppings,  long range interactions and random local fields. 
However, for spin chains with  long-range (deterministic) interactions in the presence of random magnetic fields\cite{pino14,SafaviNaini2019} for $\alpha\geq 1$ a power-law  increase with time, $S(t)\sim t^{1/(\alpha)}$ has been obtained. 
Note  however, that  for  large  $\alpha \gg 1  $ and for the considered time range,  this EE scaling 
 $S(t)\sim \exp (1/(\alpha)  \ln t )$
 is consistent with   our result  Eq. (\ref{Sdyn})  $S(t)\sim (1/\alpha)  \ln t.$ 

For $\alpha<\alpha_c$ the fixed-point of the coupling distribution is unknown. However,  the excited eigenstates of  this  model were found to follow  a sub-volumic law, with an algebraic growth of their EE with subsystem size $n$\cite{Mohdeb2022}. This is due to the existence of localized regions (dimers)\cite{Mohdeb2022}, which prevent the excited states EE to satisfy a volume law. Repeating the argumentation above,  the half-chain EE at large time $t$ is then expected to  scale as $S(t) \sim t^{\gamma(\alpha)}$, where $\gamma(\alpha) \leq 1$
is a decreasing function of $\alpha <\alpha_c$. The saturation value for the half-chain EE, is thereby expected to scale as \begin{equation} \label{slimit}
S(t\rightarrow\infty)= \ln2 \left(N/2\right)^{\gamma(\alpha)},
\end{equation}
where $N$ is the system size.\\
\begin{it}
Exact diagonalization.—
\end{it}
To check the validity of these analytical  results, we use numerical exact diagonalization and examine the half-chain EE dynamics  after a quench starting from a Néel state
$|\psi_0\rangle=|\uparrow\downarrow\uparrow\downarrow\uparrow...\rangle$.
Results are shown in Fig. \ref{DynEE} for different values of $\alpha$. We consider two different system sizes,  $N=12$, and $N=16$ to account for finite size effects. The density of spins is fixed to $n=0.1$, and averaging was done  over up to $1000$ disorder realizations. 
We see that for $\alpha>1.8$ the EE for large times shows a logarithmic enhancement with time as it was obtained via RSRG-t. The prefactor is found to be a decaying function of $\alpha$, consistent with our analytical prediction Eq. (\ref{Sdyn}). For $\alpha=1.9, 2, 3, 4$ Eq. (\ref{Sdyn}) is in  good agreement with  ED. Our approximation is expected to become more precise with increasing $\alpha \gg 1,$ where the corrections to the SDFP become smaller. 
For $\alpha\leq 1.8$ EE is found to saturate  quickly. 
The saturation occurs  faster for smaller $\alpha$. For transient times it 
grows  faster than logarithmically, as a power-law $S(t) \sim t^{\gamma(\alpha)}$, where $\gamma(\alpha)$ is a decreasing function of $\alpha\leq\alpha_c$. The
dependence of the saturation value $S(t\rightarrow\infty)$  on $N$ is in quantitative agreement  with  the scaling obtained by  RSRG-X formula\cite{Mohdeb2022},
Eq. (\ref{slimit}),   see  Fig. \ref{DynEE}. Remarkably, we note that for $\alpha=\alpha_c\approx 1.8$, the exponent $\gamma(\alpha)=0.34$ is similar to the observed universal exponent for MBL systems with long-range interactions at criticality\cite{Deng2020}. 

\begin{it}
Conclusion
\end{it}.— 
The eigenstates of  long-range interacting XY spin chains with positional disorder are known to be marginally localized when the interactions fall-off  faster than a critical power $\alpha_c \approx1$ both in the ground state\cite{Mohdeb2020} and  in the middle of the energy band\cite{Mohdeb2022}. Here, by  extending the  strong disorder renormalization group  to the quench dynamics,   we find that 
the entanglement entropy  grows with time after a  quench starting from a high-energy nonentangled state for $\alpha> \alpha_c$, logarithmically with a prefactor which is inversely proportional to $\alpha$, Eq. (\ref{Sdyn}). For  $\alpha > \alpha_c \approx 1.8$ we find good agreement with 
 the results obtained by ED. 
 This  logarithmic scaling with time differs from the  one observed in  long-range model with deterministic couplings and random on-site magnetic fields\cite{pino14,SafaviNaini2019},
 where power law time dependence was found,
 but for  $\alpha \gg 1$ both  scalings are concordant. The obtained growth of EE with time is faster than in 
 the short-range XX spin chain with random bonds, where the EE grows as a double logarithm, $S(t)\sim \ln\ln(t)$.
The faster  logarithmic growth of EE after a  quench  on long-range interacting spin systems which we found here for  exclusive bond randomness and in absence of magnetic fields could be a characteristic of quantum critical glasses with long range interactions.

For $\alpha<\alpha_c$ delocalized eigenstates were found to exist previously \cite{Mohdeb2022},
yielding to a power law increase of the EE with subsystem size. 
Building on these results we
obtained here analytical results for the large time saturation value of the half-chain EE, Eq. (\ref{slimit}), which scales with  the number of spins $N$ as $(N/2)^{\gamma}$ with $0<\gamma(\alpha)<1$ a decreasing function of $\alpha$, in good agreement with numerical exact diagonalization results.

Our analysis raises questions about the current comprehension of MBL, its critical counterpart, and their dynamical properties in presence of long-range interactions.
While the EE scaling with subsystem size and  energy level spacing statistics 
have been used widely as insightful diagnostics for MBL and criticality \cite{Luitz2015,Oganesyan2007,Huang2014,Vasseur2015,Schiffer2019},  the entanglement dynamics is found to yield  remarkably different results for  short-range\cite{bardarson12}, and  long-range interacting random spin  systems\cite{pino14,SafaviNaini2019}. Exploring the phase diagram of long-range interacting spin chains subject to  both random on-site potentials and  random bonds would help provide a more complete picture of the dynamics of quantum entanglement.

Recent advances in experimental 
setups allow to study XX spins  with interactions that fall-off as $1/r^3,$
which has been demonstrated by coupling  Rydberg states with opposite parity \cite{ex1,Signoles2021,ex3}.    Within this setup and exploiting the  particle fluctuation and correlation technique\cite{ex4,ex5} its non-equilibrium dynamics could thus be studied for $\alpha=3.0$. Chains of trapped ions with power-law interactions, decaying as $1/r^\alpha$, with tunable $0<\alpha<1.5$  have already been realized\cite{ex6,ex7,ex8} and may thereby open an experimental route  to detect the quantum phase transition transition between logarithmic and power law growth of entanglement entropy.

{\it Acknowledgements.-}
We acknowledge funding from DFG KE-807/22.


\bibliography{ref}

\end{document}